\documentclass[runningheads]{llncs}
\usepackage{float}
 
\usepackage[final,year=2024,ID=7826]{eccv}

\usepackage{eccvabbrv}
\usepackage{xspace}
\usepackage{listings}
\usepackage{caption}
\usepackage{gensymb}

\usepackage{graphicx}
\usepackage{booktabs}
\usepackage{lipsum}

\usepackage[accsupp]{axessibility}  

\newcommand{\staticTask}{\textsc{PnP}}
\newcommand{\sitHuman}{\textsc{S$_{hum}$}}
\newcommand{\sitObj}{\textsc{S$_{obj}$}}

\newcommand{\Reasoner}{\textsc{Reasoner}}
\newcommand{\Prompter}{\textsc{Prompter}}

\newcommand{\ignore}[1]{}

\usepackage[pagebackref,breaklinks,colorlinks]{hyperref}

\usepackage{orcidlink}
\usepackage[width=122mm,left=12mm,paperwidth=146mm,height=193mm,top=12mm,paperheight=217mm]{geometry}
\usepackage{graphicx}
\usepackage{comment}
\usepackage{amsmath,amssymb} 
\usepackage{color}
\usepackage{booktabs}
\usepackage{cite}
\usepackage[belowskip=-15pt,aboveskip=5pt,font=small,labelfont=bf]{caption}
\usepackage{multirow}

\usepackage{colortbl} 
\usepackage{xcolor} 
\usepackage{wrapfig} 

\usepackage{pifont} 

\newcommand{\xmark}{\ding{55}}

\definecolor{blue}{RGB}{0,0,255}
\definecolor{green}{RGB}{81,157,69}
\definecolor{red}{RGB}{255,0,0}
 
\newcommand{\greencmark}{\textcolor{green}{\ding{51}}} 

\definecolor{Gray}{gray}{0.90}
\newcolumntype{a}{>{\columncolor{Gray}}r}

\definecolor{lightblue}{rgb}{0.8,0.85,1} 
\definecolor{lightgreen}{RGB}{200, 255, 200} 

\newcommand{\dataset}{\texttt{\textbf{SIF}}}

\definecolor{existingtask}{RGB}{156, 187, 205}
\definecolor{sittask}{RGB}{255, 176, 135}

\lstdefinestyle{mystyle}{
    backgroundcolor=\color{gray!10}, 
    basicstyle=\ttfamily\footnotesize, 
    breakatwhitespace=false,
    breaklines=true,
    breakindent=0pt,
    keepspaces=true,
    numbers=left,
    numbersep=5pt,
    showspaces=false,
    showstringspaces=false,
    showtabs=false,
    tabsize=2,
    xleftmargin=0pt,
    framexleftmargin=0pt
}

\begin{document}

\title{Situated Instruction Following}


\titlerunning{Situated Instruction Following}


\author{So Yeon Min\inst{1} \and
Xavi Puig\inst{2} \and
Devendra Singh Chaplot\inst{2} \and 
Tsung-Yen Yang\inst{2} \and
Akshara Rai\inst{2} \and
Priyam Parashar\inst{2} \and 
Ruslan Salakhutdinov\inst{1} \and \\
Yonatan Bisk\inst{1} \inst{2} \and 
Roozbeh Mottaghi\inst{2} }

\authorrunning{Min et al.}

\institute{Carnegie Mellon University, Pittsburgh PA 15213, USA \and
Fair, Meta., Menlo Park CA 94025, USA}

\maketitle

\begin{abstract}
  Language is never spoken in a vacuum. It is expressed, comprehended, and contextualized within the holistic backdrop of the speaker's history, actions, and environment.
  Since humans are used to communicating efficiently with situated language, the practicality of robotic assistants hinge on their ability to understand and act upon implicit and situated instructions.
  In traditional instruction following paradigms, the agent acts alone in an empty house, leading to language use that is both simplified and artificially ``complete.''  In contrast, we propose \textit{situated instruction following} (\dataset), which embraces the inherent underspecification and ambiguity of real-world communication with the physical presence of a human speaker. The meaning of situated instructions naturally unfold through the past actions and the expected future behaviors of the human involved. 
  Specifically, within our settings we have instructions that  (1) are ambiguously specified, (2) have temporally evolving intent, (3) can be interpreted more precisely with the agent's dynamic actions.
    Our experiments indicate that state-of-the-art Embodied Instruction Following (EIF) models lack holistic understanding of situated human intention. \\
    Project website: \url{https://soyeonm.github.io/SIF_webpage/}
\end{abstract}

\section{Introduction}
\label{sec:intro}

Humans naturally engage in communication that is contextually situated, providing just enough information as necessary. This is because our use of language is predicated on an assumed common ground~\cite{Findings-of-ACL:Chandu2021}, which encompasses our shared history, actions, and environment. For instance, the instruction ``Can you bring me a cup?'' can vary in meaning depending on the context. If spoken while the speaker is donning rubber gloves by the kitchen sink, it likely refers to a dirty cup located on the living room table. Conversely, the same request made in front of the bathroom sink typically implies a need for a clean cup. While it is possible to seek clarification, humans generally interpret and respond to such requests accurately without additional information. This capability demonstrates how humans skillfully use environmental and action cues to interpret ambiguous language, crafting meanings that are intricately nuanced and context-specific.

\begin{figure*}[!t]
    \centering
    \includegraphics[width=1\linewidth]{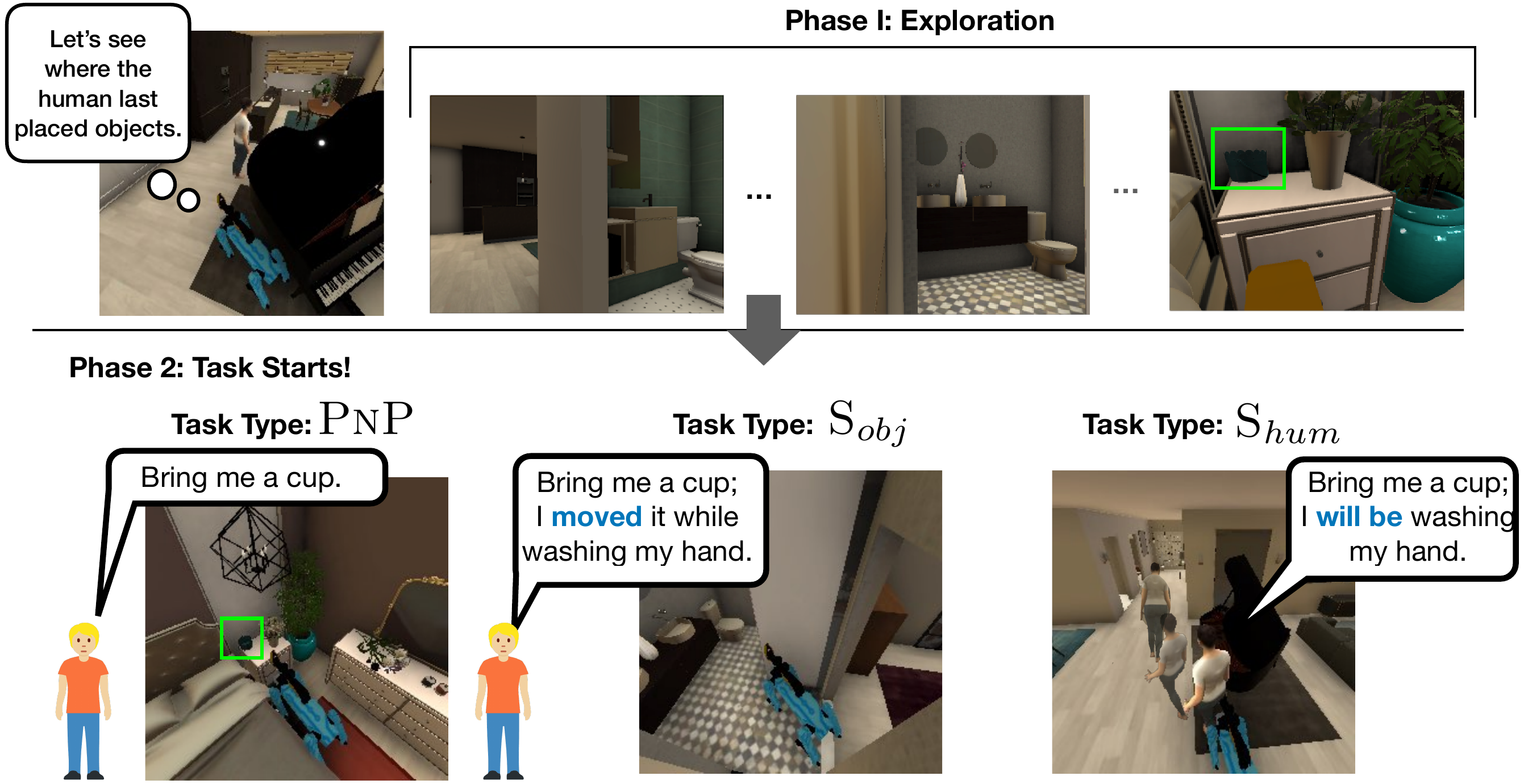}
    \vspace{-1.5em}
    \caption{\textbf{Situated Instruction Following.} The tasks in \dataset{} consist of two phases: an exploration phase (phase 1) and a task phase (phase 2). \staticTask{} represents a conventional static Pick-and-Place task used for comparison, wherein the environment remains unchanged after the exploration phase. \sitHuman{} and \sitObj{} introduce two novel types of situated instruction following tasks. In these tasks, the \textit{objects} and \textit{human} subjects move during the task phase. Nuanced communication regarding these movements is provided, necessitating reasoning about ambiguous, temporally evolving, and dynamic human intent.
    }
    \label{fig:overview}
\end{figure*}

As robotic agents increasingly become integral to our daily lives, their effectiveness and utility critically depend on their ability to comprehend and respond to situated language— natural language spoken by humans. Without this capability, agents may prove more of a hindrance than a help, forcing users to perform tasks themselves rather than entrusting them to an assistant. As discussed in the field of agent alignment \cite{leike2018scalable}, it is often difficult for users to precisely define or articulate ideal task specifications. Consequently, an agent that demands detailed explanations might render manual task execution by humans more attractive.

Current instruction-following tasks prioritize accurate low-level instruction interpretation~\cite{shridhar2020alfred,anderson2018vision,TEACH,vlnsurvey} or use commonsense to achieve underspecified goals like object navigation~\cite{chaplot2020object, das2018embodied}.
In contrast, our work \dataset{} aims to generalize \textit{Embodied} Instruction Following to \textit{Situated} Instruction Following, with instructions closer to the language naturally spoken by humans. Specifically, we focus on three dimensions of situated reasoning: 

\begin{enumerate}
    \item \textbf{Ambiguity:} As in the cup example above, there is ambiguity in the instruction given by the speaker. 
    \item \textbf{Temporal:} A speaker's actions  change how their instruction should be interpreted (e.g., clarifying an underspecified reference).
    \item \textbf{Dynamic:} When the environment changes, the agent needs to decide what actions will reduce their uncertainty (e.g., following the human). 
\end{enumerate}

We implement our tasks in Habitat 3.0\cite{puig2023habitat} as it includes simulated human agents.
To ensure a fair comparison to prior work we include tasks where the environment is static (prior work) and dynamic (this work) -- Fig.~\ref{fig:overview}. The static task is a classic pick-and-place (\staticTask) paradigm. Formally, the instructor tells the agent to \texttt{Put [Obj] in/on [Recep]}.  While prior work extends this paradigm with linguistic complexity~\cite{shridhar2020alfred, weihs2021visual, TEACH}, the baselines used for such tasks can be fairly evaluated here.  We introduce a simplifying assumption and allow the agent to explore the environment to minimize the role of mapping on the performance in our reasoning benchmark -- we will return to this in the ablations.

The focus of our benchmark is dynamic tasks where the agent must combine their understanding of the instruction with the human's movements to determine the correct action. The two domains (Figure~\ref{fig:overview}) are either where the human has moved the object (\sitObj) or where the moving human is the receptacle (\sitHuman). 
In these tasks, the agent receives a goal instruction (e.g., ``Bring me a mug'' for \sitHuman\,or ``Put the mug in the bathroom'' for \sitObj), accompanied by hints on relocation (e.g., ``I will be taking a bath'' for \sitHuman{} or ``I have moved the cup next to the be'' for \sitObj). Furthermore, in \sitHuman\,tasks, the human begins to walk at the start of the task, embedding the intent of relocation in both verbal directives and observable movement. The goal of the agent is to adhere to the instructions with an efficient path, accurately identifying and retrieving the specified object, and then placing it in the correct receptacle (which is ``human'' for \sitHuman{} tasks).
Note, these are incredibly simple tasks that will prove very difficult for current models. If someone asks you to retrieve a mug from the kitchen and bring it to the bedroom, you intuitively wait just long enough to see which room they enter before searching for the mug. These types of ambiguity reduction strategies will elude the models, exposing their over-reliance on commonsense. Moreover, if informed that a mug has been moved to a bedroom, you would thoroughly search each bedroom, ensuring no area is overlooked. Our findings show that reasoning in ambiguous situations often confuses models, posing more challenges than simple commonsense reasoning.

We specifically target evaluation of 
state-of-the-art Embodied Instruction Following (EIF) baselines. We implement two such systems inspired by papers on related tasks. The first baseline, which we refer to as \Reasoner{}, is a closed-loop system incorporating a semantic map, a prompt generator, and a Large Language Model (LLM) planner. For the prompt generator, we integrated components from Voyager\cite{wang2023voyager}, LLMPlanner\cite{song2023llm}, and ReAct\cite{react}, tailoring them to suit our dataset's specific requirements. The second baseline, \Prompter{}\cite{inoue2022prompter}, was very successful at executing ALFRED~\cite{shridhar2020alfred} tasks despite being open-loop. 
We see the desired result that our static scenarios match those from existing EIF datasets \cite{song2023llm, inoue2022prompter}, and these LLM based approaches perform very well in tasks requiring common sense. However, their performance significantly declines when faced with situations that require reasoning about the human's behavior.

\section{Related Work}
\label{sec:related_work}

\dataset{} builds on instruction following, agent alignment, and situated reasoning. 
\noindent \textbf{Instruction Following.}
We contextualize our baselines models and benchmark, in terms of both text-only and embodied research. The recent trend is to factor the task into a planner (often an LLM), memory (e.g., a map), perception, and implement tools for taking actions. Papers then often decide which components to be learned or implemented heuristically~\cite{llmagentsurvey}. 

\begin{table}[!t]
  \centering
  \label{tab:dataset_comparison}
  \scriptsize
    \caption{\small Comparison of Embodied Instruction Following Datasets.}
    \vspace{-1em}
  \begin{tabular}{lccccccc}
    \toprule
    & \multicolumn{3}{c}{\textbf{Task Specification}} & \multicolumn{2}{c}{\textbf{Env. Dynamics}} & \multicolumn{2}{c}{\textbf{Intent Change}}  \\
    \cmidrule(lr){2-4} \cmidrule(lr){5-6}  \cmidrule(lr){7-8} 
    & Abstract & Over- &  Middle- & Ego & Ego + &  Language & Intent  \\
    &  & specified &  ground  & & + Human  &  Change & Unfolding     \\
    \midrule
    ObjectNav \cite{anderson2018evaluation} & \greencmark & \xmark & \xmark  &\greencmark  & \xmark & \xmark &  \xmark \\
    ImageNav \cite{zhu2017target, krantz2022instance} &  \greencmark & \xmark &  \xmark  &\greencmark  & \xmark &  \xmark &  \xmark\\
    Course-grained VLN \cite{qi2020reverie, vlnsurvey} & \greencmark & \xmark & \xmark & \greencmark  & \xmark &\xmark &  \xmark \\
    Fine-grained VLN \cite{ku2020room, anderson2018vision} & \xmark  & \greencmark  & \xmark  & \greencmark  &  \xmark & \xmark &  \xmark \\
    Embodied QA \cite{das2018embodied} & \greencmark   & \greencmark   & \xmark   & \greencmark  &  \xmark & \xmark &  \xmark \\
    Excalibur \cite{zhu2023excalibur} & \xmark & \xmark & \greencmark & \greencmark & \xmark  & \xmark &  \xmark \\
    ALFRED \cite{shridhar2020alfred} & \greencmark & \greencmark & \xmark & \greencmark  & \xmark & \xmark  &  \xmark  \\
    TEACh \cite{TEACH}, DIALFRED \cite{gao2022dialfred} & \greencmark &  \greencmark &  \xmark  & \greencmark & \greencmark & \greencmark & \xmark  \\
    \midrule
     \rowcolor{lightgreen} \dataset{} & \greencmark & \greencmark & \greencmark & \greencmark &\greencmark  &\greencmark  &\greencmark \\
    \bottomrule
  \end{tabular}
  \label{tab:comparison}
  \vspace{-2em}
\end{table}

\label{sec:eif}
\paragraph{Embodied Instruction Following (EIF)} \ignore{EIF tasks are typically divided into three categories: semantic navigation, manipulation, and mobile manipulation. }We focus on tasks involving high-level task planning and navigation, excluding those with low-level (joints and end effector) motion control \cite{rb2, nist, toto}. Relevant tasks are summarized in Table~\ref{tab:comparison}.

Modular models with the structure of planner (e.g. LLM), memory (e.g. semantic mapping), and perception/execution tools have shown much success in the tasks of our interest. In semantic navigation, ESC \cite{esc} and l3vm\cite{yu2023l3mvn} have achieved SOTA performance in ObjectNav\cite{chaplot2020object,min2023self, anderson2018evaluation} and ImageNav\cite{imagenav, zhu2017target, krantz2022instance} by leveraging LLMs for frontier selection based on common sense. \cite{zhou2023navgpt, chen2024mapgpt, shah2023lm} have used LLM for planning and tracking in Visual Language Navigation \cite{vlnsurvey}. In mobile manipulation, FILM\cite{min2021film, min2022don}, \Prompter{} \cite{inoue2022prompter}, and LLMPlanner\cite{song2023llm} have used language models for planning and semantic search in ALFRED\cite{shridhar2020alfred}. In real-robot domains, code-as policies \cite{liang2023code}, inner monologue \cite{huang2022inner}, and ProgPrompt\cite{singh2023progprompt} have shown success in using LLMs for generating abstracted manipulation plans.

The effectiveness of decoupled LLM reasoners largely stems from existing benchmarks focusing on abstract common-sense reasoning \cite{shridhar2020alfred, chaplot2020object, das2018embodied, qi2020reverie}  or detailed, sequential guidance \cite{vlnsurvey, anderson2018vision, ku2020room, TEACH}. These modular models tackle visual grounding and navigation through separate semantic memory modules \cite{esc, min2021film, chaplot2020object, shah2023lm}. Thus, the \textit{learning} challenges posed by either abstract instructions or detailed guidance boil down to common-sense object search and visually grounded tracking. Yet, these foundational skills fall short of meeting real-world needs and utility. Real-world instructions often exist in a middle ground: not overly abstract yet not excessively detailed, with both explicit statements and implied expectations. Thus, we introduce \dataset{}, aiming to surpass the limitations of basic reasoning and literal planning by bridging the gap to real-world applicability.

\paragraph{Text-only agents} Modularly chained models of planner\cite{gpt3point5, achiam2023gpt, touvron2023llama}, memory (in text) \cite{wu2023plan}, meta-reasoner\cite{yao2024tree}, and refiner\cite{madaan2024self} have shown much success in code generation\cite{zhang2024codeagent}, playing games\cite{wu2024spring}, web navigation\cite{zhou2023webarena}, and text-only embodied agents\cite{react}. While most efforts have concentrated on enhancing agent reasoning through tool-use \cite{react, schick2024toolformer} and self-evaluation (reflection\cite{madaan2024self, shinn2024reflexion} and tree-of-thought \cite{yao2024tree}), less attention has been paid to navigating the complexities of real-world instructions. \cite{li2023eliciting, zhang2023clarify, hla} have explored eliciting clarifications when human intent is ambiguous. Clarification is necessary at times, but it also comes at a cost; the focus of \dataset{} is understanding ambiguous intent that is clear upon holistic understanding of the environment and speaker. \cite{hla, effenberger2021analysis, suhr2019executing} have studied when instruction intent changes and is communicated with language; \dataset{} instead infers changes to intent from the human's actions. 
Unlike prior work focusing on agent-centric challenges in dynamic settings (e.g. moving zombies)\cite{wu2024spring, wang2023voyager}, \dataset{} explores understanding dynamically evolving human intent.

\label{sec:related_inst}

\noindent\textbf{Agent Alignment.}
Traditional agent alignment research\cite{leike2018scalable, agarwal2019learning, malik2021inverse} recognizes the difficulty in articulating real-world task specifications, and has primarily focused on learning reward functions from underspecified specification. Recent works on LLM and LLM agents\cite{wang2023voyager, wu2024spring, zhou2023webarena} have delved into alignment for real-world applications, focusing on understanding intent\cite{ouyang2022training}, value alignment with social norms \cite{evolving} and individual preference \cite{yang2024towards}, and guaranteeing safety\cite{tian2023evil}. Despite these advancements, the ability of agents to interpret instructions characterized by ambiguity, evolving intent, and the interplay between agent actions and environmental dynamics is still underexplored. We introduce \dataset{}, aiming to push the boundaries of agent alignment closer to real-world complexity.

\noindent\textbf{Situated Reasoning.}
Environments evolve when there is another agent other than oneself; reasoning about action and change requires situated logical reasoning. While early work \cite{mccarthy1963situations, reiter1991frame} and STAR\cite{wu2021star} employed formal logic and scene-hypergraphs for abstract reasoning, \dataset{} focuses on fostering holistic understanding and temporal awareness in uncertain contexts.

\ignore{
\begin{figure*}[t]
    \centering
    \includegraphics[width=1\linewidth]{figures/desired_behavior.png}
    \caption{\textbf{Clear/nuanced tasks example and desired model behavior}: The goal of our dataset is to. To do so, it consists of 2 phases and three different taks types. }
    \label{fig:desired_behavior}
\end{figure*}}

\section{Dataset}

\dataset{} extends previous work that primarily focus on abstract, decontextualized, common-sense reasoning \cite{gao2022dialfred, chaplot2020object, TEACH} by evaluating reasoning scenarios constructed in situ. Below, we explain the dataset design choices and guiding philosophy.

\subsection{Tasks}
\label{sec:taskexp}

\noindent \textbf{Overview.} Our tasks are structured into two distinct phases: (1) the exploration phase and (2) the task phase. During the exploration phase, the agent is allotted $N$ steps to navigate around a static house environment where object assets are positioned. The value of $N$ is determined to ensure the agent has sufficient steps to thoroughly scan the environment; specifically, $N = $ 1.5 x (the number of steps required to achieve a complete map using frontier-based exploration techniques). Following the exploration phase, some objects are repositioned without the agent's knowledge. As the task phase commences, the agent receives an instruction (e.g., ``Bring me a cup,'' ``Put the cup in the sink''), accompanied by either direct or ambiguous information regarding which objects have been moved (e.g., ``I took a cup with me. I'll be getting ready for bed''). If the task involves delivering an object to a human, the human walks into the agent's field of view as the task begins, simultaneously providing hints about their intended location (``I will be in the bathroom washing my face''). These elements, along with other strategic design decisions, ensure that the exploration phase effectively contextualizes the language directives, rendering tasks sufficiently solvable.

\begin{wraptable}{r}{0.44\linewidth}
\vspace{-28pt}
\centering
\footnotesize

\caption{\textbf{\small Dataset stats}. 
Data statistics and seen/unseen splits are balanced across conditions.}
\vspace{1.5em}
\begin{tabular}{lccc}
   \textbf{Axis} & \textbf{\staticTask} & \textbf{\sitObj} & \textbf{\sitHuman} \\
\toprule
\textbf{Dynamic}& &  &  \\ 
\; Object
                & \xmark & \greencmark & \xmark  \\    %
\; Human
                & \xmark & \xmark & \greencmark  \\    %
\midrule 
\midrule 
\textbf{\# Tasks} & & & \\ 
\; Valid \tiny{Seen/Unseen}
                & \scriptsize{40/40} & \scriptsize{40/40} & \scriptsize{40/40}   \\    
\; Test \tiny{Seen/Unseen}
                & \scriptsize{40/40} & \scriptsize{40/40}  & \scriptsize{40/40} \\    
\bottomrule
\end{tabular}
\label{tab:stats}
\vspace{-2em}
\end{wraptable}

\noindent \textbf{Task Format.} A task is defined by $\langle H, I, C, P_e, P_t, P_g\rangle$. A task is embedded in the $H$\textit{ouse}, with starting $P$\textit{oses} of assets during , agent, and the human for the  $e$\textit{xploration} ($P_e$), $t$\textit{ask} ($P_t$) phases and $g$oal state ($P_g$), the goal $I$\textit{nstruction}, and humans $C$\textit{ommunication} about which objects were moved where after exploration phase. 

\noindent \textbf{Agent Specs \& Predefined Skills.} 
The agent in our simulation is modeled after the Spot robot\cite{BostonDynamicsSpot}, equipped with an odometry sensor and RGB and Depth cameras mounted on its arm at a height of 0.73 meters; this setup follows the specifications of 640x480 resolution and a 79-degree horizontal field of view by previous work~\cite{chaplot2020object}. At each timestep, the agent is capable of executing one of several actions: moving forward by 0.17 meters, rotating left or right by 10 degrees, grabbing an object, or placing an object. The ``grab object'' action allows the agent to pick up the nearest graspable object within a 2-meter radius, whereas the ``put object'' action enables it to place the held item onto the nearest suitable receptacle within the same distance.

\ignore{As detailed further in Section 4, we also introduce high-level skills such as "Go to Room $X$," "Explore Room $X$," and "follow the human," underpinned by semantic mapping. This framework not only facilitates a more intuitive interaction between the agent and its environment but also provides dataset users with a versatile toolkit. They can choose to leverage the predefined tools and skills we have developed or opt for a more direct approach by employing end-to-end style models such as Seq2Seq\cite{shridhar2020alfred}, Palm-E\cite{driess2023palm}, RT-X\cite{padalkar2023open}, using language commands alongside RGB and Depth inputs for navigation and task execution.}

\noindent \textbf{Task Types.}  
There are three types of tasks - static (\staticTask), situated-object (\sitObj), and situated-human (\sitHuman). For all three types, the overarching goal for the agent is to pick \texttt{[Obj]} and place it in/on \texttt{[Recep]}. In the \sitObj{} tasks, the \texttt{[Obj]} of interest is relocated between the exploration and task phases; in the \sitHuman{} tasks, the \texttt{[Recep]} (human) begins to move at the start of the task phase. In the \staticTask{} task type, both \texttt{[OBJ]} and \texttt{[RECEP]} remain stationary; although other objects may be moved to introduce variability, the \texttt{[OBJ]} relevant to the instruction ($I$) is not displaced. This setup is analogous to tasks found in existing research \cite{shridhar2020alfred, chaplot2020object} and serves as a means for sanity checking and ensuring a fair comparison across models. Examples of \staticTask, \sitObj{}, \sitHuman{} tasks are in Fig.~\ref{fig:overview}b.

\paragraph{Three dimensions of situated reasoning}  We explain how the aspects of ambiguous, temporal, and dynamic are implemented below:

\begin{itemize}
    \item \textbf{Ambiguous}: Ambiguity in \sitObj\,tasks arises when multiple potential locations exist for searching an object, as informed by communication cues. For instance, if a human says, ``I took the cup and moved it with me. I am washing my face,'' but there are multiple bathrooms in the house, the task becomes ambiguous. Likewise, ambiguous \sitHuman\ tasks arise when the human could potentially be in several different locations. The ambiguity in these tasks was annotated by human reviewers. 
    \item \textbf{Temporal}: For example, if a human says, ``Bring me the cup. I will be in the bathroom,'' and there are multiple bathrooms, the intent (the exact location of which bathroom) unfolds and becomes more clear real-time, with the human walking towards one of the bathrooms (Fig.~\ref{fig:human_traj}).
    \item \textbf{Dynamic}: In both \sitObj{} and \sitHuman{} tasks, an agent can decide  which location to search or whether to follow the human or decide, to decrease ambiguity or understand temporally unfolding intent. 
\end{itemize}

\paragraph{Two axes of difficulty} Our tasks incorporate commonsense related to room functions and human activities, based on object placement (Tables ~\ref{tab:template},\ref{tab:com_template}). We add two layers of difficulty to this foundation: 

\noindent \textbf{Holistic understanding of language instructions}: 
We designed and filtered tasks to avoid being trivial or solvable solely through commonsense, yet not so ambiguous as to require exhaustive search. Phase 1 exists so that the agent can scan the layout of the house, and tasks are solvable with targeted reasoning rather than comprehensive searching.\\[0pt]

\noindent \textbf{Resolving ambiguity}: Tasks necessitate methodical reasoning under ambiguous intent (ambiguous tasks of \sitObj{} and \sitHuman{}), with designs that favor agents taking actions to resolve ambiguity (intent probing), rather than comprehensively searching later. At the same time, unambiguous tasks should be solved without intent probing. Evaluation details in Sec.~\ref{sec:results}.

\begin{table}[!t]
\begin{minipage}[!t]{.45\textwidth}
\footnotesize
\caption{Template Instructions ($I$)}
\vspace{-1em}
\begin{tabular}{@{}l@{}}
\toprule
\textbf{Task Descriptions} \\
\midrule
\textbf{Static Receptacle:}\\
Put a \texttt{[ObjectCat]} on the \texttt{[Recep]} \\
Put a \texttt{[ObjectCat]} in the \texttt{[RoomFunction]} \\
\textbf{Dynamic Receptacle:} \\
Bring a \texttt{[ObjectCat]} to me \\
\bottomrule
\end{tabular}
\label{tab:template}
\end{minipage}
\hspace{2em}
\begin{minipage}[!t]{.45\textwidth}
\centering
\footnotesize
\caption{Template Relocations ($C$)}
\vspace{-1em}
\begin{tabular}{@{}l@{}}
\toprule
\textbf{I moved the object with me; I am ...} \\
\midrule
next to the \texttt{[Recep]}\\
next to the \texttt{[Recep1]}, \texttt{[Recep2]}, \texttt{[Recep3]} \\
in \texttt{[RoomFunction]} \\
in \texttt{[RoomFunction][WithObjects]} \\
doing [\texttt{RoomFunctionActivity]} \\
\bottomrule
\end{tabular}
\label{tab:com_template}
\end{minipage}
\vspace{-2em}
\end{table}

\subsection{Dataset Construction}

We explain how the tuple $\langle H, I, C, P_e, P_t, P_g\rangle$ is constructed. 
In a total of 10 houses, each with human-annotated room metadata (which includes details such as the top-down (x,y) coordinates corresponding to each room, the function of the room (e.g., bedroom), and grounding details (e.g., a bedroom with a yellow wall)), we place four to ten assets in \texttt{[ObjectCat]}\footnote{basket, book, bowl, cup, hat, plate, shoe, stuffed\_toy}. Assets are from YCB\cite{calli2017yale}, Google Scanned Objects\cite{downs2022google}, and ABO \cite{collins2022abo} datasets.
We sample $P_e, P_t$ of assets so that they are initialized in a visible space and graspable. More details on filtering trivial and unsolvable tasks are in Appendix~\ref{sec:filtering}.\\

\noindent\textbf{Language Directives.} We explain the generation of $I$ and $C$. The $I$nstruction carries information about $P_g$, which is the desired location of an object necessary for task success. We first sample $P_g$, by sampling a room in the scene, a receptacle in the room, and then a puttable point on the receptacle. Then, we use templated language (Table~\ref{tab:template}) to express this information. For \staticTask{} and \sitObj{} tasks, we use ``Put a \texttt{[ObjectCat]} on the \texttt{[Recep]}'' or ``Put a \texttt{[ObjectCat]} in the \texttt{[RoomFunction]}''; for \sitHuman{} tasks, we use ``Bring a \texttt{[ObjectCat]} to me''. The list of possible \texttt{[ObjectCat]}, \texttt{[Recep]}, \texttt{[RoomFunction]} is shown in Table~\ref{tab:complete_args}. 

The $C$ommunication on movement describes $P_t$ of objects whose poses are changed from $P_e$. For each object that was moved, $C$ is given with the templates in Table~\ref{tab:com_template}. In addition, for \sitHuman\,types, $C$ also contains where the human will be after walking during the task phase; this also follows the template of Table~\ref{tab:com_template}, with ``I am'' replaced by ``I will be'' (e.g., $I$: `Bring a cup to me'. $C$: `I will be organizing my bed'). \texttt{[RoomFunctionActivity]} describes common human activities in each room (e.g., kitchen - washing vegetables, bedroom - preparing to sleep); the complete list is presented in Table~\ref{tab:roomfunction}. 
$C$ is presented together with $I$ at the start of the task phase. Although we use templated language, the diverse combination with the scene layout and object/human poses introduces substantial reasoning challenges.\\

\noindent\textbf{Human Trajectories.} For \staticTask{} and \sitObj{} tasks, the human is stationary. 
 In \sitHuman{} tasks, the human trajectory is deterministically determined given the human's $P_t$ and $P_s$. The human moves naturally at a speed of 0.08m per time step. Human appearance and motion is naturally implemented as explained in \cite{puig2023habitat}. \\

\noindent\textbf{Statistics and Splits} The specifics of the dataset are outlined in Table~\ref{tab:stats}. For validation and testing, the dataset comprises 40 seen and 40 unseen tasks across each type, resulting in 240 validation and 240 testing tasks in total. Additionally, we provide code to facilitate further data and trajectory generation for training purposes. The seen subsets (both validation and test) incorporate the same six houses used in the training, allowing for evaluation in familiar environments. Conversely, the unseen subsets employ four new houses to test generalization across different settings, bringing the total to ten unique houses for evaluation. 

\section{Baselines}
\label{sec:baseline}

As discussed in Sec.~\ref{sec:related_work}, many recent state-of-the-art EIF agents are modular models with an LLM planner, connected to learned/engineered episodic memory, perception, and execution tools. \ignore{More specifically, these agents use semantic mapping as episodic memory (constructed from depth/semantic segmentation perception modules), learned/algorithmic navigation and manipulation tools, and a text-representation generator that converts the episodic memory into text. }We present two baselines within this high-performing family. The first is \Reasoner{}, a closed-loop baseline that adapts FILM\cite{min2021film} and the prompts of llm-planner\cite{song2023llm}, and ReAct \cite{react}, and  prompter \cite{inoue2022prompter}, an open-loop SOTA agent built for ALFRED \cite{shridhar2020alfred}. 

\begin{figure*}[t]
    \centering
    \includegraphics[width=1.0\linewidth]{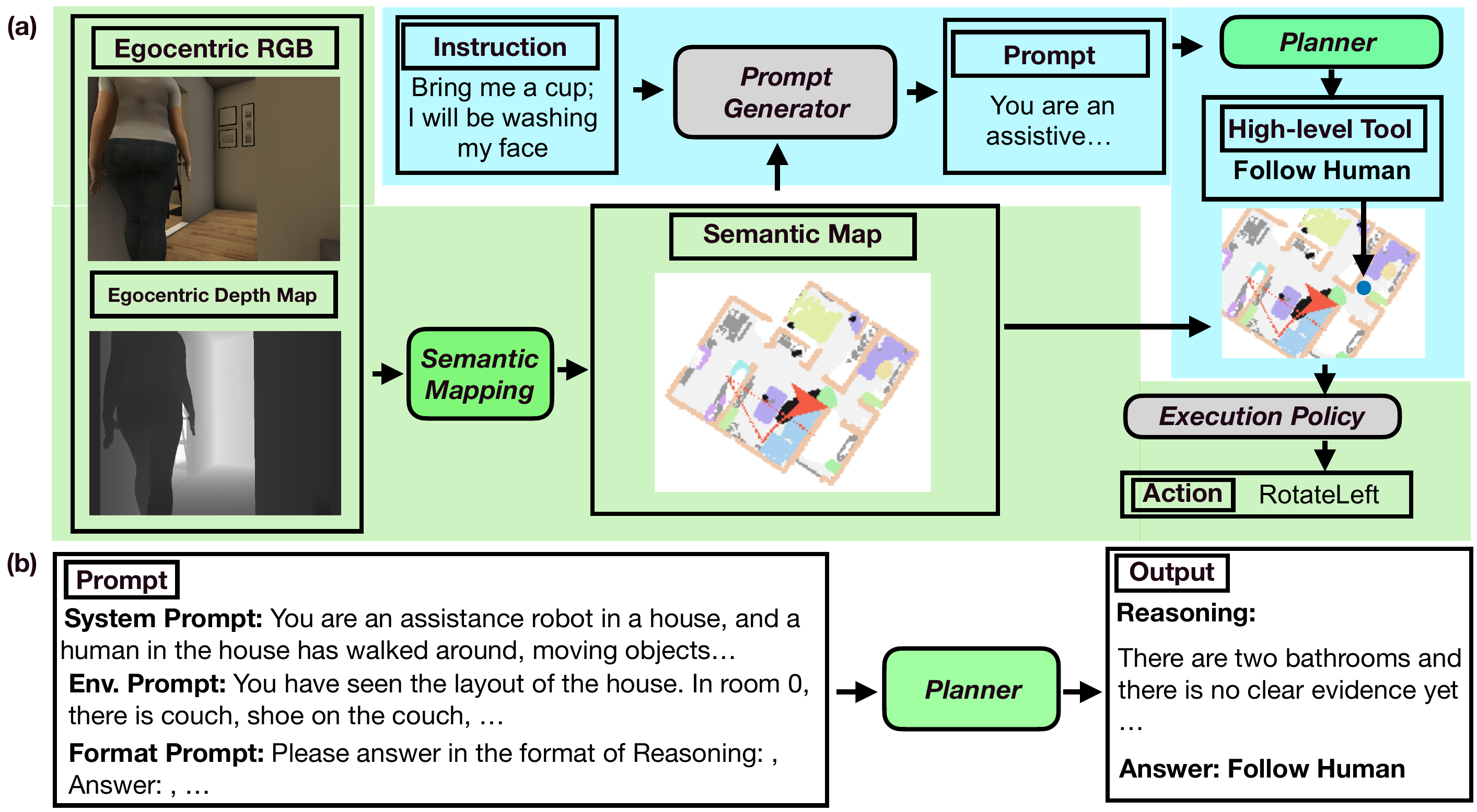}
    \caption{\small \textbf{\Reasoner{}}: (a) The semantic mapper is updated at every timestep, whereas the prompt generator and planner are activated either upon completion of the last high-level action or when a new decision is required. (b) The prompt consists of system prompt, environment prompt, format prompt. }
    \label{fig:baseline}
\end{figure*}

\subsection{Reasoner}

We adopt the modular structure of FILM~\cite{min2021film}. 
\Reasoner{} operates through three main components: (1) a semantic mapper that updates an allocentric map from egocentric RGB and depth inputs, (2) a prompt generator, and (3) the planner (GPT-4o\cite{achiam2023gpt}) that generates high-level actions (Fig.~\ref{fig:baseline}). 

\paragraph{Semantic Mapper.} The semantic mapper creates a global representation for visual observation. As in previous work\cite{chaplot2020object, min2021film}, we process egocentric RGB and depth into an allocentric top-down map of obstacles and semantic categories using Detic\cite{zhou2022detecting}. The semantic categories of interest are \texttt{[ObjectCat]}, \texttt{[Recep]}, and ``human.'' In contrast to previous works\cite{chaplot2020object, min2021film}, the most recent human and object positions are refreshed post new observations and pick/place actions, ensuring a dynamic and accurate representation of the environment (Sec.~\ref{sec:map_update}).

\begin{figure*}[t]
    \centering
    \includegraphics[width=0.98\linewidth]{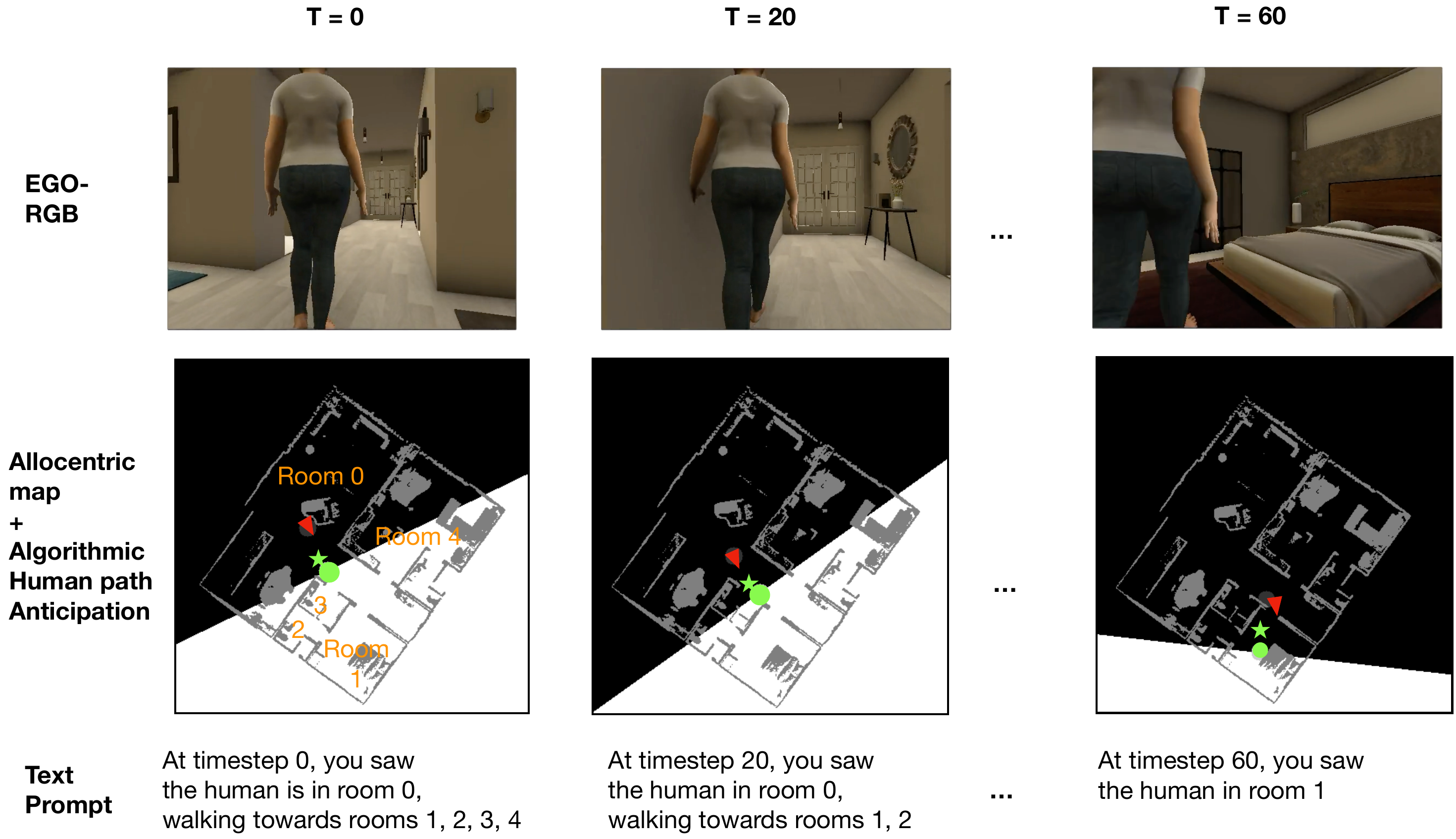}
    \caption{\small \textbf{Text Prompt Generation of Human Trajectory}: The white regions in the maps are possible regions that the human might walk towards; rooms with more than half of the area included in the white region are included in the text prompt. The red triangle is the agent position/direction, green star and dot are respectively current observed human position, anticipated human position in 10 steps. The text prompt at every 20 timesteps is given to \Reasoner{} (and at time step 0 to \Prompter{} which is open-loop), to decide if there is enough evidence for the clarity of the human's intent. }
    \label{fig:human_traj}
\end{figure*}

\paragraph{Text representation generator.} The semantic map and other contexts are converted into prompts. It is a concatenation of three components: the system prompt, environment prompt, and the format prompt:
\begin{itemize}
    \item \textbf{System:} The system prompt outlines the agent's role and and encourages it to account for uncertainty. It is presented as ``You are an assistive robot in a house, aiding a human. Your observations may be incomplete or wrong."
    \item \textbf{Environment:} The environment prompt is a conversion of the episodic memory into text format, and contains information of the agent's current state and previously completed/failed actions. It is given in the following sequence: (1) observation of $P_e$ during exploration phase, based on the semantic map, (2) $C$, regarding object/ human movements, (3) the goal instruction $I$, (4) the high-level action executed by agents at timesteps and their observed consequences (success/fail), (5) the agent's latest observation, based on the latest semantic map (example in Appendix~\ref{sec:prompt}). Every observation is given with the caveat that it can be incorrect or missing. Past actions are interleaved with observations as in ReAct.

    \item \textbf{Format:} The format prompt explains action affordance and a format for chain of thought \cite{wei2022chain}. 
    It also explains the desired effect of actions (e.g. ``If you want to keep searching for object(s) or human that might exist (but you have not detected) in the current room, choose `Explore Room $X$' (Table~\ref{tab:tools}).'')

\end{itemize}

A complete example of the prompt is given in Appendix~\ref{sec:prompt}. For \textit{\sitHuman} tasks, every 20 steps, we ask whether the planner has enough evidence about the human's goal location, based on the system prompt, the current observation of human trajectory (Fig.~\ref{fig:human_traj}), and the human's utterance (e.g. ``I will be organizing my bed''). Example prompts about the observed human trajectory are in Fig.~\ref{fig:human_traj}. If the planner answers ``Enough Evidence,'' we proceed to ask for a high-level action (e.g. ``Grab [Obj]''), providing the concatenation of the system prompt, environment prompt, and the format prompt above. On the other hand, if it answers ``Not Enough Evidence,'' we call ``follow human.'' We ask this question every 20 steps, until the ``follow human'' execution tool deems that either the agent lost track of the human or the human has stopped. Details are in Appendix~\ref{sec:human_follow}. 

\begin{wraptable}{r}{0.48\textwidth}
\centering
\small 
\vspace{-0.75em}
\caption{\small Execution tools for \Reasoner{}; working details/affordance in Tab.~\ref{tab:tools}.}
\vspace{1.5em}
\begin{tabular}{@{}l@{\hspace{10pt}}l@{}}
\toprule
\textbf{Navigation} & \textbf{Manipulation} \\
\midrule
Go to Room $X$ &  Grab Obj \\
Explore Room $X$   & Put Obj \\
Follow Human & Give Obj to Human. \\
\bottomrule
\end{tabular}
\label{tab:tools_shortened}
\vspace{-5em}
\end{wraptable}

\vspace{-1.5em}
\noindent \paragraph{Execution Tools} Upon receiving the prompt, the planner is prompted to choose a high-level action (Tab.~\ref{tab:tools_shortened}); then corresponding execution tools are called. A complete list of tools are listed in Table~\ref{tab:tools}. When the execution is done, the tool sends this message, and  the prompt generator creates a new prompt and the planner calls a new tool. 

\subsection{\Prompter{}}

\Prompter{} employs an open-loop planner. We utilize GPT 3.5 for planning and search in lieu of BERT\cite{devlin2018bert} (as in the original work\cite{inoue2022prompter}), tailored to our dataset's requirements. Its operational logic is straightforward: if an object has been identified on the map, \Prompter{} interacts with it; otherwise, it initiates a search based on object and receptacle relationships. The planning phase utilizes a template that is populated with specific details, executed once at the task's outset. For instance, in response to a command like ``Put a shoe on the couch," GPT 3.5 is prompted to formulate a high-level plan adhering to a structured format (e.g. ``\texttt{[Pick up OBJ, Put on the RECEP]})." This ensures GPT 3.5 primarily focuses on filling in the template's variables. For \sitHuman{} tasks, we ask if there is enough evidence about the human goal location, based on the layout of the house and the human utterance; due to open-loop planning, we ask this only at the beginning of the task. If the planner answers ``Not enough evidence'', then we add ``Follow Human'' as the first high-level action (so the plan becomes e.g. \texttt{[Follow Human, Pick up OBJ, Give OBJ to Human]}).

For the search phase, \Prompter{}, in the original work, determines where to look by sampling from the logit values from a text query akin to ``Something you find at [MASK] is apple.'' Instead of sampling directly, we query the LLM, by integrating the latest map's text representation into a search prompt (e.g., ``In which room is the shoe likely to be? Please answer in the format of Room X.''). This caters to our dataset's complexity, featuring multiple smaller rooms. Since object search happens multiple times, this has the same effect as sampling.\footnote{We grid-search the LLM's hyperparameters — temperature and top\_p, by trying temperature $\in \{0.1, 0.3, 0.5\}$ and top\_p $\in \{0.1, 0.3, 0.5\}$ in the validation set.}

\begin{table}[!t]
  \centering
  \footnotesize
       \caption{\small \textbf{SPL} performance of \Prompter{} and \Reasoner{} across splits. In each sectioned-row, the top row assumes oracle perception (semantic segmentation and manipulation); the bottom row assumes learned semantic segmentation and heuristic manipulation. To minimize the burden on API costs and time, we have limited LLM API calls for plan generation to 15 times for both \Prompter{} and \Reasoner{}. SR performance is shown in Table~\ref{tab:results_sr}.}
    \vspace{-1em}
    \begin{tabular}{@{}llrrr@{\hspace{1em}}rrr@{\hspace{2em}}rrr@{\hspace{1em}}aaa@{}}
    \toprule
    \multicolumn{2}{l}{Model}& \multicolumn{3}{c}{Val Seen} & \multicolumn{3}{c}{Val Unseen} &  \multicolumn{3}{c}{Test Seen} & \multicolumn{3}{c}{Test Unseen} \\
    \cmidrule(lr){3-5} \cmidrule(lr){6-8} \cmidrule(lr){9-11} \cmidrule(lr){12-14} 
    Planning & Perception 
    & \rotatebox{90}{\staticTask} & \rotatebox{90}{\sitObj} & \rotatebox{90}{\sitHuman} &  \rotatebox{90}{\staticTask} & \rotatebox{90}{\sitObj} & \rotatebox{90}{\sitHuman} & \rotatebox{90}{\staticTask} & \rotatebox{90}{\sitObj} & \rotatebox{90}{\sitHuman} &  \rotatebox{90}{\staticTask} & \rotatebox{90}{\sitObj} & \rotatebox{90}{\sitHuman} \\
    \midrule
    Oracle & Oracle  & 100 & 100 & 98& 100 & 100 & 100	& 98 & 100 & 95 & 98 & 100 & 100 \\
           & Learned & 47 & 44 & 59 & 41 & 35 & 53 & 49 & 27 & 71 & 36 & 47 & 55  \\
    \midrule
    \textbf{\Prompter{}}\cite{inoue2022prompter} 
    & Oracle  & 70 & 29 & 29 & 67 & 32 & 23	& 65 & 26 & 26 & 56 & 32 & 25 \\
    & Learned & 17 & 10 & 11 & 22 & 8 & 16 & 27 & 3 & 14 & 19 & 7 & 13\\
    \midrule
    \textbf{\Reasoner{}} 
    & Oracle &  81 & 60 & 29 & 77 & 49 & 36	& 74 & 59 & 29& 81 & 52 & 38\\
    & Learned & 19 & 11 & 15 & 25 &	10 & 13	& 29 & 3 & 19 & 20 & 16 & 13\\
    \bottomrule
    \end{tabular}
\label{tab:results}
\vspace{-2em}
\end{table}%

\vspace{-0.3cm}
\section{Results and Analysis}
\vspace{-0.3cm}
\label{sec:results}

We evaluate the baseline models against the validation and test splits of \dataset.
\vspace{-0.2cm}
\paragraph{Oracle:} The oracle baseline replaces the learned planner of \Reasoner{} with ground-truth plans like [``Go to Room $X$'', ``Grab Obj $X$'']. Its purpose is to demonstrate the most reasonable path length achievable with optimal reasoning strategies. Further information is in Appendix~\ref{sec:oracle_baseline}.
\vspace{-0.2cm}
\paragraph{Evaluation Metrics:} Task Success is determined by whether the object is correctly placed in or on the intended receptacle—or the right room—within 600 timesteps. A clear task in \sitHuman\ is deemed unsuccessful if it necessitates following the human for over 50 timesteps, since unconditional probing should not is a behavior that reduces the utility of the robot. The Success Rate is calculated as the average of individual task successes across the dataset. The SPL (Success weighted by Path Length) \cite{anderson2018evaluation} is calculated using the formula:

$$ SPL = E_{tasks}[s \frac{L^{*}}{max(L, L^*)}],  $$
where $s$ is task success, $L$ is path-length outputted by the model for a task, and $L^*$ is the path given by the oracle baseline.
SPL is the primary metric of evaluation, because it tests correct reasoning strategies. 

\vspace{-0.5cm}
\subsection{Results}
\vspace{-0.2cm}

Results from our experiments are presented in Table~\ref{tab:results}. This table notably shows the following facts about our dataset and baselines. First, the gap of model performance (both \Reasoner{} and \Prompter{}) across \staticTask{} versus \sitHuman, \sitObj{} shows that \staticTask{} can be solved with commonsense and mechanistic combination, and the rest two tasks cannot. \Prompter{} shows SPL of $\sim$20\% ($\sim60$\% with oracle perception) on \staticTask{} tasks, showing that these tasks are on par with existing tasks like ALFRED. Conversely, on \sitObj{} and \sitHuman{} tasks, it shows much lower performance ($\sim30$\% with reasoning alone); this shows that these tasks have challenges beyond common sense and progress tracking. The reasoning challenges of \sitObj{} and \sitHuman{} are backed by the performance of \Reasoner{} with oracle perception/manipulation; it shows a stark contrast in \staticTask{} tasks ($\sim75$\%) and \sitObj, \sitHuman{} tasks ($\sim50$\%). Overall, \Reasoner{} shows a better performance than \Prompter{}, with and without GT perception/manipulation.

Second, it shows that perception is still a bottleneck, as found in works on previous datasets\cite{min2021film, min2022don, chaplot2020object}. Even the oracle baseline, which has perfect reasoning, suffers with learned semantic segmentation and manipulation. When combined with learned reasoning, the drop tends to be larger (oracle perception vs. learned perception of \Prompter{} and \Reasoner{}), since the planner faces a further uncertainty from perception error.

\vspace{-0.5cm}
\subsection{Ablations and Analysis}

\begin{wraptable}[9]{r}{.43\textwidth}
\vspace{-5pt}
\centering
\footnotesize
\captionof{table}{\small Ablation SR with Oracle plan, for visual and execution errors on Val Seen \& Unseen combined.}
\vspace{1.3em}
\begin{tabular}{@{}lccc@{}}

\toprule
\textbf{Method} & {\textbf{\staticTask}} & {\textbf{\sitObj}} & {\textbf{\sitHuman}} \\
\midrule
G.T. Oracle & 100 & 100 & 99 \\
+ heuristic man. & 83& 78 &88 \\
+ learned seg. & 64 & 54 & 78 \\
+ both & 49 & 49 & 63\\
\bottomrule
\end{tabular}
\label{tab:ablations}
\end{wraptable}

To analyze the impact of non-reasoning components (visual perception, manipulation) on task performance, Table~\ref{tab:ablations} shows the effect of using heuristic manipulation and learned segmentation across our three settings. Our findings align with previous studies~\cite{min2021film, inoue2022prompter, min2023self}, further emphasizing that segmentation continues to be a significant obstacle to progress.

\begin{wraptable}[14]{r}{.43\textwidth}
\centering
\footnotesize
\vspace{-25pt}
\small \captionof{table}{Reasoning Error Modes. Percentage of failed tasks for each error (w/ oracle perception) on Val Seen \& Unseen combined.}
\vspace{1.5em}
\renewcommand{\arraystretch}{0.8}
\begin{tabular}{l@{\hspace{1em}}cccccccc}
\toprule
\multirow{2}{*}{\textbf{Error}} & \multicolumn{3}{c}{\Reasoner{}} && \multicolumn{3}{c}{\Prompter{}} \\
\cmidrule{2-4} \cmidrule{6-8}
&  \rotatebox{90}{\textbf{\staticTask}} & \rotatebox{90}{\textbf{\sitObj}} & \rotatebox{90}{\textbf{\sitHuman}} &&  \rotatebox{90}{\textbf{\staticTask}} &\rotatebox{90}{\textbf{\sitObj}} & \rotatebox{90}{\textbf{\sitHuman}} \\
\midrule
Parsing \hspace{1em}& - & 5 & -  &&  46 & 30 & -  \\
Planning & 22 & 15 & - && -  & - & - \\
Strategy & 44 & 80 & 100 && 46 & 43 &  100 \\
~ \scriptsize  Obj & - & 50 & - && - & 33 & -  \\
~ \scriptsize Room & 44 & 30 & - && 46 & 27 & -  \\
~ \scriptsize Human & - & - & 100 && - & - &  100 \\
Actuation & 33 & - & -  && 8 & 9 & - \\
\bottomrule
\end{tabular}
\label{tab:error_modes}
\vspace{-100pt}
\end{wraptable}

The focal point of our study is reasoning. To this end, Table~\ref{tab:error_modes} analyzes the failure modes of \Reasoner{} and \Prompter{} when they leverage accurate semantic segmentation and manipulation. This table enumerates the proportions of various reasoning error modes that we observed in unsuccessful tasks. It categorizes these errors as follows: parsing errors, where the format of the LLM's response deviates from expectations; planning errors, which include inadequate tracking of progress and incorrect actions; strategic errors in locating humans, objects, or rooms; and actuation errors, such as mismanaging object interactions. The table lists these errors chronologically, noting that early errors like parsing mistakes can preclude later ones. Notably, strategic errors often manifest as unnecessary repetitive movements—more than five high-level navigational or exploratory actions—or as overly confident yet inaccurate predictions about a human's location.

This table reveals several key insights. First, \staticTask{} tasks exhibit fewer strategy errors compared to the other two tasks, which happens primarily due to room grounding issues with less obvious rooms when placing objects---for instance, identifying a ``study room.'' This implies that \staticTask{} tasks revolve around more common-sense reasoning, such as inferring room functions from objects and familiar human activities typically associated with those rooms. In contrast, \sitObj{} and \sitHuman{} tasks exhibit higher rates of strategy errors due to their need for a more situated understanding of human motion and activity. Second, \Prompter{}'s open-loop planning leaves it more susceptible to parsing errors. Notably, \Prompter{} often fails to generate a plan for instructions that, while atypical, are feasible (e.g., ``Put a shoe on the couch''), responding with ``I'm sorry, but I can't comply with that request.'' In contrast, \Reasoner{} exhibits fewer fundamental errors (those that precede strategy errors) and records a higher count of strategy errors. However, this does not imply that \Reasoner{} commits more strategic mistakes than \Prompter{}; from Table~\ref{tab:results}, it achieves a higher number of successful tasks with correct strategic execution.

\begin{table}[!t]
\centering
\setlength\tabcolsep{1pt} 
\scriptsize
\captionof{table}{\small \textbf{Ambiguous vs Clear tasks}. SPL and SR of \Reasoner{} and \Prompter{} with G.T./learned vision and manipulation on Val seen \& unseen combined.}
\vspace{-1em}
\begin{tabular}{llccccccccccc}
\toprule
\textbf{Model} & \textbf{Metric} & \multicolumn{5}{c}{\textbf{G.T. Vis. \& Man.}} & & \multicolumn{5}{c}{\textbf{Learned Vis. \& Man.}} \\
\cmidrule{3-7} \cmidrule{9-13} 
& & \multicolumn{2}{c}{\textbf{\sitObj}} && \multicolumn{2}{c}{\textbf{\sitHuman}} && \multicolumn{2}{c}{\textbf{\sitObj}} && \multicolumn{2}{c}{\textbf{\sitHuman}} \\
\cmidrule{3-4} \cmidrule{6-7} \cmidrule{9-10} \cmidrule{12-13} 
&& Clear & Amb. && Clear & Amb. && Clear & Amb. && Clear & Amb. \\
\midrule
\multirow{2}{*}{\Reasoner{}} 
& SPL & 58 & 55	&& 8 & 47 && 8 & 12 && 2 &22 \\ 
& SR & 72 & 77 && 15 & 75 && 15 & 18 && 3 & 33 \\ 
\midrule
\multirow{2}{*}{\Prompter{}}& SPL 
 & 41 & 24 && 4 & 41	&& 11 & 8 && 0 & 22  \\ 
& SR 
 & 58 & 31 && 4 & 65 && 18 & 10 && 0	& 30\\ 
\bottomrule
\end{tabular}
\label{tab:ambiguity}
\vspace{-2em}
\end{table}

Table~\ref{tab:ambiguity} examines model performance on clear versus ambiguous tasks. Ambiguity in \sitObj{} tasks emerges when multiple potential locations exist for an object, as indicated by communicative cues (Sec.~\ref{sec:taskexp}). For example, the statement ``I am washing my face'' becomes ambiguous when multiple bathrooms are available. Similarly, ambiguous \sitHuman{} tasks occur when the human could be in several different locations. Ambiguous \sitObj{} tasks require a systematic search of possible locations; common failures include models fixating on the incorrect room across multiple attempts or missing a potential room altogether. Clear and ambiguous \sitHuman{} tasks also demand careful interpretation of the ambiguity of instructions (e.g., deciding whether to follow the human), informed by both contextual clues and observations of the environment and human movement (Fig.~\ref{fig:human_traj}).

For \sitObj{} tasks, \Reasoner{} exhibits consistent performance across both clear and ambiguous conditions, demonstrating its capability to navigate between rooms and conduct systematic searches, whereas \Prompter{} displays significant performance discrepancies, struggling particularly with ambiguous scenarios. In \sitHuman{} tasks, both models underperform in clear tasks due to a tendency to conservatively judge that there is insufficient evidence of the human's destination, even when only one plausible location exists. \Prompter{} notably has an almost zero success rate in clear tasks, consistently concluding insufficient evidence and choosing to follow the human. Conversely, \Reasoner{} attempts some calibration but generally leans towards following the human. Qualitative analysis reveals that in ambiguous tasks, \Reasoner{} often disengages prematurely, assuming it has accumulated enough evidence.

\section{Conclusion}
\vspace{-0.5em}
We present Situated Instruction Following (\dataset{}), a new dataset to evaluate situated and holistic understanding of language instructions. Our dataset reflects aspects of real-world instruction following: (1) ambiguous task specification, (2) evolving intent over time, and (3) dynamic interpretation influenced by agent action. \dataset{} is carefully crafted to assess language comprehension and reasoning in situ. We show that current state-of-the-art models struggle with this level of understanding, further highlighting the complexity and uniqueness of our dataset.

\par\vfill\par
\clearpage  

%
%
\bibliographystyle{splncs04}
\bibliography{egbib}
\clearpage 

\appendix

\section{Task Details}

\subsection{Task Filtering}
\label{sec:filtering}
Tasks that are invalid or trivial are filtered, using the oracle agent (Section~\ref{sec:results}). First, a task is invalid if, for the goal (\texttt{[Obj]}) asset, its $P_e$ is not findable or $P_t$ is not reachable. To filter tasks with invalid $P_e$, we check whether \texttt{[Obj]} was detected in the oracle agent's semantic map during the exploration phase; to filter tasks with invalid $P_t$, we run the oracle agent for task phase, and filter tasks where ``Grab Obj'' was unsuccessful. Furthermore, a task is trivial if, for the goal (\texttt{[Obj]}) asset, one of $<P_e, P_t>$ or $<P_t, P_g>$ are very close or within the same receptacle. This makes the task trivial because there is not much change between exploration and task phase, or between the initial state at task phase and the goal state. We use simulator data to access the parent receptacle of \texttt{[Obj]} asset for $P_e, P_t, P_g$ and filter the task if any of these belong to the same parent receptacle.

\subsection{Details on Language Directives}

\setcounter{table}{0} 
\renewcommand{\thetable}{\Alph{section}.\arabic{table}}

The complete list of arguments (Sec.~\ref{sec:taskexp}) for language directives are shown in Tab.\ref{tab:complete_args} The full list of [RoomFunctionActivity] is shown in Table~\ref{tab:roomfunction}.

\begin{table}[ht]
\centering
\caption{Complete list of arguments (\texttt{[ObjectCat]}, \texttt{[Recep]},  \texttt{[RoomFunction]}) for language directives $I$ and $C$ (Sec.~\ref{sec:taskexp})}
\begin{tabular}{l@{\hspace{2em}}l@{\hspace{2em}}l }
\toprule
\texttt{[ObjectCat]} & \texttt{[Recep]} & \texttt{[RoomFunction]} \\
\midrule
basket & chair &  living room \\
book   & shelves & bedroom  \\
bowl      & bed & kitchen  \\
cup & toilet & dining room  \\
hat  & bench & bathroom  \\
plate    & bathtub & garage \\
shoe     & couch & empty room  \\
stuffed\_toy  & counter &  dressing room\\
  & table &  study room  \\
\bottomrule
\end{tabular}
\label{tab:complete_args}
\end{table}

\begin{table}[!h]
\centering
\caption{Human Activity, by each room function}
\begin{tabular}{l@{\hspace{2em}}l}
\toprule
Room & Activities \\
\midrule
living room & watching TV; hanging out near the couch; vaccuming the living room \\
bedroom    & preparing to sleep;  organizing my bed; reading on my bed \\
kitchen    & washing vegetables; preparing my meal; sorting groceries \\
dining room & setting up the table; eating dinner \\
bathroom   & washing my face; washing my hand; taking a bath; \\
& brushing my teeth; shaving \\
garage     & washing my car; fixing my car \\
empty room  & meditating in the empty room; stretching in the exercise room \\
dressing room & choosing on my outfit; trying on clothes; organizing my clothes \\
study room & studying; cleaning my desk \\
\bottomrule
\end{tabular}
\label{tab:roomfunction}
\end{table}

\section{Prompt Examples}
\label{sec:prompt}

\subsection{\Reasoner{} prompt examples}

We show an example prompt for a \sitObj{} task (Prompt 1). 
Furthermore, we show an example ambiguity calibration prompt for a \sitHuman{} task (Prompt 2).

\label{sec:reasoner_prompt}
\begin{lstlisting}[breaklines=true,caption={\textbf{Prompt 1:} Example prompt for \Reasoner{} for a \sitObj{} task at timestep 125.}]
% System Prompt
You are an assistive robot in a house, helping a human. Your observations may be incomplete or wrong.

While premapping, you saw the layout of the house;  you saw that Room 0 has: chair, shoe on a chair, couch, book on a couch, table; Room 1 has: fridge, counter, cabinet; Room 2 has: chair, bed, table. Room 3 has: chair, stuffed_toy on a chair, bench, table, stuffed_toy on a table. Room 4 has: toilet. Your observations (and the map) can be stale/ incorrect/ missing, because the human moved objects since you had premapped the house, or your vision system is imperfect.

Then, a little after that, the human said `I took a shoe and moved it with me. I put it in the livingroom'. A little after that, the human said `I took a book and moved it with me. I put it in the livingroom'. A little after that, the human said `I took a stuffed_toy and moved it with me. I put it in the livingroom'. A little after that, the human said `I put a book in the livingroom'. A little after that, the human said `I moved a shoe. I put it in the livingroom'. After all of this, the human finally commanded to you: 'Put a book in bedroom.' "

% Environment Prompt
Timestep 12 : You further explored another location in room 2. Based on the preamap and the parts of the rooms that you scanned just now, in map 12 (at time step 12), you saw that Room 2 has: chair, bed, table; Your observations (and the map) can be stale/ incorrect/ missing, because the human moved objects since you had premapped the house, or your vision system is imperfect.
Timestep 24 : You further explored another location in room 2. Based on the preamap and the parts of the rooms that you scanned just now, in map 24 (at time step 24), you saw that Room 2 has: chair, bed, table; Your observations (and the map) can be stale/ incorrect/ missing, because the human moved objects since you had premapped the house, or your vision system is imperfect.
Timestep 125 : You checked room 0; you scanned some parts of it (and did not scan other parts). Based on the preamap and the parts of the rooms that you scanned just now, in map 125 (at time step 125), you saw that Room 0 has: chair, shoe on a chair, couch, book on a couch, table, shoe on a table; Your observations (and the map) can be stale/ incorrect/ missing, because the human moved objects since you had premapped the house, or your vision system is imperfect.

Combining your latest observations, In map 125 (at time step 125), you saw that Room 0 has: chair, shoe on a chair, couch, book on a couch, table, shoe on a table; Room 1 has: fridge, counter, cabinet, human; Room 2 has: chair, bed, table. Room 3 has: chair, stuffed_toy on a chair, bench, table, stuffed_toy on a table. Room 4 has: toilet. Your observations (and the map) can be stale/ incorrect/ missing, because the human moved objects since you had premapped the house, or your vision system is imperfect.

% Format Prompt
The human said your goal is to ``Put a book in bedroom''. The human may be nuanced or ambiguous. To achieve the goal, choose the next action among Available Actions: [`Explore Room 0', `Go to Room 1', `Go to Room 2', `Go to Room 3', `Go to Room 4', `Pick up shoe', `Pick up book', `Pick up shoe', `Done!']' .

You are in room 0 and you can ONLY grab/ put objects inside Room 0. If you want to keep searching for object(s) or human that might exist (but you have not detected) in the current room, choose `Explore Room 0'. To go to another room, choose ``Go to Room 'another room number'''.
 Please ONLY respond in the format: RESPONSE FORMAT: Reasoning: reason about the next action. Action: one of Available Actions (e.g.Explore Room 0). If task is complete, answer Action: Done!
 Now, which action should you take? RESPONSE FORMAT:
\end{lstlisting}

\begin{lstlisting}[breaklines=true,caption={\textbf{Prompt 2:} Example ambiguity calibration prompt for \Reasoner{} for a \sitHuman{} task at timestep 103.}]
% System Prompt
You are an assistive robot in a house, helping a human. Your observations may be incomplete or wrong.
While premapping, you saw the layout of the house;  you saw that Room 0 has: chair, stuffed_toy on a chair, bed, stuffed_toy on a bed, bench, couch, stuffed_toy on a couch, table; Room 1 has: chair, bed, couch, table, cabinet; Room 2 has: chair, bathtub, table, cabinet; Room 3 has: bed, fridge, cabinet. Room 4 has: chair, bed, fridge, table, cabinet. Room 5 has: chair, toilet, plate on a toilet, couch, book on some receptacle , bowl on some receptacle . Room 6 has: bed, toilet, fridge, cabinet. Room 7 has: toilet, bathtub, cabinet. Your observations (and the map) can be stale/ incorrect/ missing, because the human moved objects since you had premapped the house, or your vision system is imperfect.
Then,  After all of this, the human finally commanded to you: `Bring a shoe to me. I will be washing vegetables.'  Right after the human said this, you saw the human in room 1, starting to walk towards somewhere. When you last saw the human, the human was in room 4, walking towards rooms 0, 2, 3, 6, 7.

% Environment Prompt
Timestep 2 : You were following the human;the human was in room 1.

Timestep 22 : You were following the human;the human was in room 1.

Timestep 42 : You were following the human;the human was in room 1.

Timestep 62 : You were following the human;the human was in room 1.

Timestep 82 : You were following the human;the human was in room 4, walking towards rooms 0, 2, 3, 4, 6, 7.

Timestep 102 : You were following the human;the human was in room 4, walking towards rooms 0, 2, 3, 6, 7.

% Format Prompt
Combining your latest observations, In map 102 (at time step 102), you saw that Room 0 has: chair, book on a chair, stuffed_toy on a chair, bed, book on a bed, stuffed_toy on a bed, bench, couch, stuffed_toy on a couch, table, book on a table; Room 1 has: chair, book on a chair, shoe on a chair, stuffed_toy on a chair, bed, shoe on a bed, couch, book on a couch, shoe on a couch, stuffed_toy on a couch, table, book on a table, shoe on a table, cabinet, book on a cabinet; Room 2 has: chair, hat on a chair, bathtub, book on a bathtub, table, cabinet, book on a cabinet, hat on a cabinet; Room 3 has: bed, fridge, book on a fridge, cabinet, book on a cabinet. Room 4 has: chair, book on a chair, bed, book on a bed, bench, fridge, couch, book on a couch, table, book on a table, cabinet; Room 5 has: chair, book on a chair, bowl on a chair, shoe on a chair, toilet, couch, book on a couch, bowl on a couch, stuffed_toy on a couch. Room 6 has: bed, toilet, fridge, book on a fridge, cabinet, book on a cabinet. Room 7 has: toilet, book on a toilet, bathtub, book on a bathtub, cabinet, book on a cabinet. Your observations (and the map) can be stale/ incorrect/ missing, because the human moved objects since you had premapped the house, or your vision system is imperfect.
The human said your goal is to ``Bring a shoe to me. I will be washing vegetables''. The human may be nuanced or ambiguous.Now, from the human's utterance (Bring a shoe to me. I will be washing vegetables) and your observations of the human (When you last saw the human, the human was in room 4, walking towards rooms 0, 2, 3, 6, 7.), do you have enough evidence which room (among rooms0, 1, 2, 3, 4, 5, 6, 7 ) the human is walking towards and will stop in?Please ONLY respond in the format: RESPONSE FORMAT: Reasoning: reason about `Enough Evidence'/ `Not Enough Evidence'. Action: choose among [`Enough Evidence', `Not Enough Evidence']. RESPONSE FORMAT:
\end{lstlisting}

\subsection{\Prompter{} prompt examples}

We show an example planning prompt (Prompt 3) and search prompt (Prompt 4) for a \sitObj{} task. Furthermore, we show ambiguity calibration prompt for a \sitHuman{} task (Prompt 5).

\begin{lstlisting}[breaklines=true,caption={\textbf{Prompt 3:} Example planning prompt for \Prompter{} for a \sitObj{} task at timestep 0.}]
The human commanded to you: `Put a hat in bathroom.' Please make a high level plan. Please answer in the format of ``[Pick up OBJ, Put in the ROOM]''. For example, ``[Pick up saltshaker, Put in the bedroom]''.
\end{lstlisting}

\begin{lstlisting}[breaklines=true,caption={\textbf{Prompt 4:} Example search prompt for \Prompter{} for a \sitObj{} task at timestep 154.}]
You are an assistive robot in a house, helping a human. Your observations may be incomplete or wrong.While premapping, you saw the layout of the house;  Room 0 has: chair, bench, couch, counter, table, ; Room 1 has: chair, bed, table, ; Room 2 has: fridge, counter, cabinet, . Room 3 has: chair, bed, . Room 4 has: chair, bed, counter, . Room 6 has: bathtub, table, cabinet, . Room 7 has: fridge, . Room 8 has: chair, toilet, cabinet. The human had said `I took a shoe and moved it with me. I put it in the kitchen'.; `I took a basket and moved it with me. I put it in the livingroom'.; `I took a stuffed_toy and moved it with me. I put it in the bedroom'.; `I took a basket and moved it with me. I put it in the livingroom'.\n In which room, is the basket likely to be in? Please answer in the format of Room X. For example, Room 1. Please ONLY respond in the format: Answer: Room X.
\end{lstlisting}

\begin{lstlisting}[breaklines=true,caption={\textbf{Prompt 5:} Example ambiguity calibration prompt for \Prompter{} in a \sitHuman{} task at time step 0.}]
You are an assistive robot in a house, helping a human. Your observations may be incomplete or wrong.While premapping, you saw the layout of the house;  Room 0 has: shelves, bathtub, fridge, couch, table, ; Room 1 has: bed, table, ; Room 2 has: shelves, fridge, cabinet, . Room 3 has: fridge, . Room 4 has: chair, shelves, table, . Room 5 has: toilet, bathtub, . The human commanded to you: `Bring a stuffed_toy to me. I will be in study room.' Right after the human said this, you saw the human in room None. Now, from the human's utterance (Bring a stuffed_toy to me. I will be in study room) and your observations of the human, do you have enough evidence which room (among rooms0, 1, 2, 3, 4, 5 ) the human is walking towards and will stop in? Please choose among [`Enough Evidence', `Not Enough Evidence'].
\end{lstlisting}

\section{Execution Details}
\subsection{Map Update}
\label{sec:map_update}
In the semantic map, position of objects and humans are algorithmically updated. There are channels for the latest position of the human and the entire human trajectory. The latest human position channel is updated to the projection from the current semantic segmentation mask, upon detecting a human. A dedicated human trajectory channel tracks all observed movements; the latest timestep is inputted to the projected area from the semantic segmentation mask of the human. For grabbable objects, updates occur upon Grab Obj/Put Obj actions. When a Grab Obj action is executed and the grasper is closed, the closest target object is removed from the map, reflecting the agent's changed perception. When Put Obj is executed and the object is detected in the agent's view, the projection of the object is put back on its corresponding channel.

\subsection{Execution Tools}
Execution tools for \Reasoner{}/\Prompter{} and their working details/affordance are in Table~\ref{tab:tools}.

\begin{table}[!h]
\centering
\begin{tabular}{lp{0.75\linewidth}}
\toprule
Execution Tool & Description \& Affordance \\
\midrule
\textbf{Navigation}\\
Go to Room X &  FMM Planner navigates to a random point in Room X. \\
Explore Room X & FMM Planner navigates to a random point in Room X; then, agent turns 15 times to the right to look around. \\
Follow Human & The last observed position of the human is given as the goal, to the human-following wrapper (more explanation is Sec.~\ref{sec:human_follow}) on top of FMM Planner. \\
\midrule
\textbf{Manipulation}\\
 Grab Obj &  The closest object within 2 meters of the grasper is grabbed, and agent's grasper is closed. \\
 Put Obj &  Grasped object is put on the closest receptacle within 2 meters of the grasper is grabbed, and agent's grasper is opened. \\
 Give Obj to Human & The agent goes within 1 meter of the human and gives grasped object to human, if human is visible from current view. \\
\bottomrule
\end{tabular}
\caption{Execution tools for \Reasoner{}/\Prompter{} and their working details/affordance.}
\label{tab:tools}
\end{table}
\subsection{Following and Anticipating the Human}
\label{sec:human_follow}
When the ``Follow Human'' skill is called (Table~\ref{tab:tools}), \Reasoner{} and \Prompter{} uses an algorithmic execution tool to follow the human. First, the latest position of the human is set as the goal, and the FMM Planner navigates until agent is within 1m. of the goal. When the goal is reached, the agent rotates 40 times to get a full 360$\degree$ view, to scan the last location of the human. If the human is visible in a new location, this location is set as the new goal for the FMM Planner. Otherwise, stop is declared; in the prompt, it is stated that ``the agent either lost track of the human or the human stopped.'' 

To describe human motion in prompts (illustrated in the bottom row of Fig.~\ref{fig:human_traj}), we use the following process. If the human has been visible for over 30 timesteps, we estimate their expected position by extending a vector from their position 30 timesteps prior to their last observed position, scaled to represent 10 future timesteps. The end point of this vector is the expected human position (the green star in the middle row of Fig.~\ref{fig:human_traj}). We then create a plane orthogonal to this vector, selecting the side opposite to the most recent human position (the depicted white side in the middle row of Fig.~\ref{fig:human_traj}). Rooms covered more than 50\% by this white area are included in the "walking towards" section of the prompt (e.g. rooms 1 and 2 at $T=20$ in Fig.~\ref{fig:human_traj}). If no room meets this criterion, "walking towards" is omitted, and the prompt only mentions the current room with the human (e.g. human seen in room 1 at $T=60$ in Fig.~\ref{fig:human_traj}).

\section{Detailed Results}
\label{sec:appendix_results}
\subsection{Oracle Baseline}
\label{sec:oracle_baseline}

The oracle baseline is established by supplying the reasoner with the ground truth plan and adding exploration as required. For \sitObj{} and \staticTask{} tasks, the protocol is to activate \texttt{explore\_room} when the target object is not immediately found. For \sitHuman{} tasks, we employ a trajectory that avoids following, as this approach is consistently more efficient than those involving following, regardless of whether the tasks are clear or ambiguous. These trajectories are considered the optimal paths for computing the SPL (Success weighted by Path Length). However, there are a small number of instances when the oracle trajectory might fail, as documented in Table~\ref{tab:results}. In these cases, we default to using the maximum permitted task duration of 600 timesteps as the optimal path length for SPL calculation.

\subsection{Full Results with Success Rate}
\label{sec:sr_table}
Results on success rate, that correspond to the SPL results of Table~\ref{tab:results} are shown in Table~\ref{tab:results_sr} below.

\begin{table}[h]
  \centering
  \footnotesize
    \begin{tabular}{llrrr@{\hspace{1em}}rrr@{\hspace{2em}}rrr@{\hspace{1em}}rrr}
    \toprule
    \multicolumn{2}{l}{Model}& \multicolumn{3}{c}{Val Seen} & \multicolumn{3}{c}{Val Unseen} &  \multicolumn{3}{c}{Test Seen} & \multicolumn{3}{c}{Test Unseen} \\
    \cmidrule(lr){3-5} \cmidrule(lr){6-8} \cmidrule(lr){9-11} \cmidrule(lr){12-14} 
    Planning & Perception 
    & \rotatebox{90}{\textbf{\staticTask}} & \rotatebox{90}{\textbf{\sitObj}} & \rotatebox{90}{\textbf{\sitHuman}} &  \rotatebox{90}{\textbf{\staticTask}} & \rotatebox{90}{\textbf{\sitObj}} & \rotatebox{90}{\textbf{\sitHuman}} & \rotatebox{90}{\textbf{\staticTask}} & \rotatebox{90}{\textbf{\sitObj}} & \rotatebox{90}{\textbf{\sitHuman}} &  \rotatebox{90}{\textbf{\staticTask}} & \rotatebox{90}{\textbf{\sitObj}} & \rotatebox{90}{\textbf{\sitHuman}} \\
    \midrule
     Oracle & Oracle  &  100 & 100 & 98 & 100 & 100 & 100 & 98 & 100 & 95 & 98 &100 &100 \\
           & Learned & 55	& 55 & 68 &	43 & 43	 & 60 & 55 & 30 & 78 & 38 & 53 & 63\\
    \midrule
    \textbf{\Prompter{}}\cite{inoue2022prompter} 
    & Oracle  & 73 & 40 & 45 & 68 & 43 & 35 & 68 & 35 & 38 & 58 & 40 & 40 \\
    & Learned & 18 & 13 & 15 & 25 & 13 & 23	& 28 & 3 & 20 & 20 & 8 & 20 \\
    \midrule
    \textbf{\Reasoner{}} 
    & Oracle & 90 & 83 & 50	& 88 & 68 & 55 & 85 & 78 & 43 & 90 & 68 & 58  \\
    & Learned & 23 & 15 & 23 & 33 & 18 & 20 & 38 & 5 & 30 & 28 & 23 & 18 \\
    \bottomrule
    \end{tabular}
    
     \caption{\textbf{SR} performance of \Prompter{} and \Reasoner{} across splits. In each sectioned-row, the top row assumes oracle perception (semantic segmentation and manipulation); the bottom row assumes learned semantic segmentation and heuristic manipulation.}
\label{tab:results_sr}
\end{table}

\end{document}